# A Comparison of Violin Bowing Pressure and Position among Expert Players and Beginners

Yurina Mizuho[1] and Yuta Sugiura[1]

[1] *Keio University, Yokohama, Japan*

(Email: ymizuho@keio.jp)

**Abstract ---** The violin is one of the most popular musical instruments, but mastering it requires a significant amount of practice time. The bowing action (pressure, position, speed) of the right hand is crucial in determining tonal quality, but this is difficult to master. This study compared the bowing movements, specifically bow pressure, bow position, and bow speed, of experienced players with those of beginners. Identifying common bowing characteristics of experienced violin players can aid the evaluation of beginners' skills and the development of a feedback system that supports self-practice and instruction.

**Keywords:** violin, bowing, motion capture

## 1 INTRODUCTION

The violin is one of the most popular string instruments. The pitch is determined by pressing the strings with the left hand, while the sound is produced by bowing the strings with the right hand. Instruments, such as the violin, viola, and cello, are called bowed string instruments. With these instruments, the bowing action has a significant impact on tonal quality. For the violin, the balance of three parameters mainly determines tonal qualify: the bow speed, distance between the bow and the bridge, and bow pressure on the string [1]. Other important elements of bowing that affect tonal quality include the bow position, angle, and tilt. To control the tonal quality, players must be able to quickly and precisely adjust these parameters. Many players adjust their playing movements based on their own experience and intuition. Therefore, achieving a wide variety of tones on the violin requires extensive practice time and experience.

Various studies have been conducted to measure bowing motions and provide feedback to support bowing practice. Some of these studies utilized motion capture to measure bow movements [2]. Others estimated bow pressure using strain gauges [3] or optical sensors [4]. Some studies used motion capture to provide haptic feedback on the correct bowing path [5], and others offered real-time feedback on sound quality and bow angle [6]. In the related studies, to verify the practice effectiveness of the feedback system for beginners, the bowing action was compared in a preliminary investigation prior to user studies. The results revealed common bowing characteristics of experienced players, which beginners do not have, and demonstrated the importance of providing feedback.

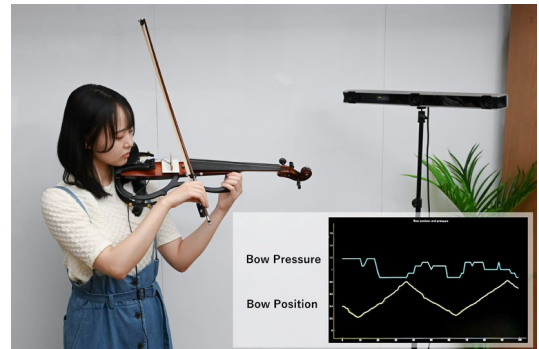

Fig.1 Bowing measurement using motion capture

By comparing the bowing actions of experienced players and beginners, common elements among experienced players can be identified. Such information can be used to evaluate beginners' skills levels and support efficient practice. Blanco et al. [6] demonstrated differences in sound stability, pitch stability, and bow trajectory between experienced players and beginners. Verrel et al. [7] focused on the cello, another bowed string instrument, and found that experienced players consistently maintained a bow-to-string angle close to 90° and exhibited smaller variations in joint angles of the elbow and wrist. These studies focused on the bow angle and the angle of the right arm. Although they also measured bow pressure and speed, they did not compare these parameters in detail.

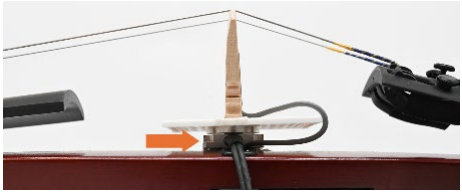

Fig.2 Load cell installed under the bridge

Therefore, this study aimed to provide a more detailed comparison of the bowing action of experienced players and beginners. Specifically, using motion capture and force sensors (Fig.1), we measured bow position, bow speed, and bow pressure. As bow pressure and speed are less visually apparent compared to bow trajectory, they require auditory evaluation, which can be challenging for beginners and those practicing independently whose ears are not well trained. By identifying bowing characteristics common to experienced players in these elements, we can contribute to efficient practice support and feedback systems.

## 2 EXPERIMENT

### 2.1 Bowing pressure measurement

We used a Yamaha Silent Violin (SV250) and a Yamaha bow (CBB101). We measured bow pressure using a load cell (TEC Gihan Co., USL06-H5-50N) installed under the bridge, as shown in Fig. 2, focusing on the z-axis force. We placed a 5 cm × 5 cm white plate on the load cell to facilitate the fixation of the bridge in a position perpendicular to the body of the violin. The voltage signal from the load cell was amplified by an amplifier (TEC Gihan Co., DSA-03A) and transmitted to a microcontroller (Arduino Pro Mini, 3.3 V). The microcontroller converted the analog signal into sensor values and calculated the force in Newtons by applying the load cell's specific calibration factor. We connected the microcontroller to a computer via USB and obtained the forces through serial communication using Python.

### 2.2 Bow position measurement

We used an optical motion capture system (Optitrack V120: Trio) to measure the bow position. We attached five markers on the violin body and five on the bow. The arrangement was based on the study by Schoonderwaldt et al. [2]. We used the software Motive to obtain the coordinates of the markers and streamed the coordinates in real-time via socket communication using Python. The bow position was calculated from the coordinates of the markers.

The bow position refers to the contact point between the bow hair and the string along the length of the bow. The bow position was numerically expressed, with the frog (near the player's hand) as 0 and the tip (farthest from the player's hand) as 1. Bow speed was calculated from the change in bow position over time using the timestamps of the acquired data.

### 2.3 Experimental protocol and participants

The experimental protocol was as follows. First, we gave beginners a brief explanation of how to hold the violin and the fundamental movements of the bow and arm. Next, we adjusted the position, height, and angle of the motion capture camera according to the participant's height. Then, we held a practice session to check the camera setup. Following this, the main session was repeated three times.

In each session, we first recorded the initial value of the load cell with the bow not touching the string. This initial value corresponded to the pressure on the bridge when the strings were tuned, and this value was subtracted from the actual measured data to record the bow pressure. Next, we set a metronome to a tempo of 75 bpm. After giving a signal, we asked the participants to perform a down-bow (from the frog to the tip) over four counts, followed by an up-bow (from the tip to the frog) over the next four counts and to continue this back-and-forth motion until giving the end signal. The participants played the open A string. We asked them to move the bow back and forth within the marked sections at the frog and tip. We acquired the data for bow pressure and position at 60 fps, with 1,500 frames captured per session.

We recruited 16 participants: eight experienced violin players (three males, five females, average age 23.4 ± 1.32 years) and eight beginners (four males, four females, average age 23.6 ± 1.93 years). The experienced players had an average of 11.9 ± 4.51 years of violin experience. Among these players, in terms of lesson experience, three had more than 15 years, three had 10 to 12 years, and three had 6 to 9 years. Among the beginners, five participants had experience with other instruments, but none had experience with bowed string instruments.

## 3 RESULTS AND DISCUSSION

Fig. 3 shows some examples of the time-series data for one trial of each participant's main session. The horizontal axis of each graph represents time, the left vertical axis represents bow pressure, and the right vertical axis represents bow position, displayed as blue and orange graphs, respectively. The data for the experienced players are shown on the left, and the data for the beginners are presented on the right. The missing data in the graph indicates that the camera failed to capture the bow movement.

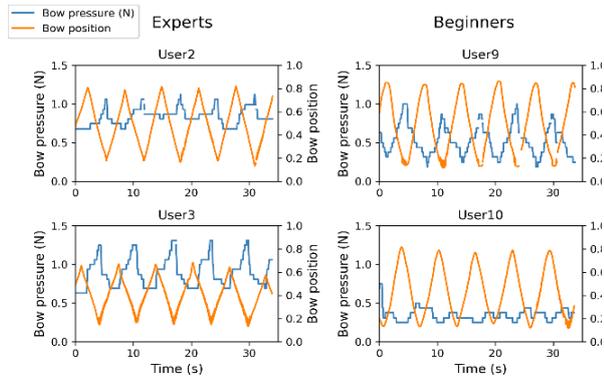

Fig.3 Time series data of bow pressure and position

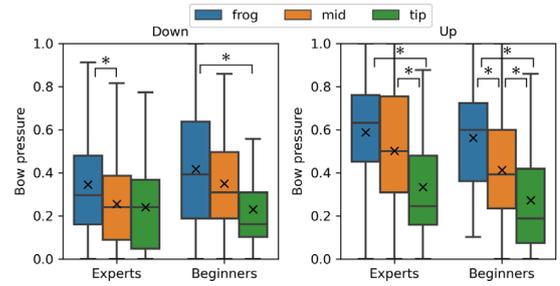

Fig.4 Relationship between bow position and pressure

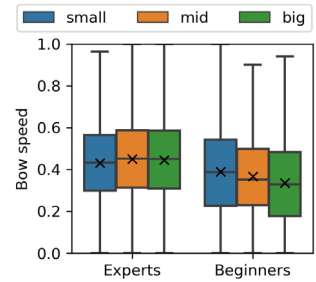

Fig.5 Relationship between bow pressure and speed

### 3.1 Bow pressure

We tested the maximum value, minimum value, and range of bow pressure using the Brunner-Munzel test at a significance level of 5%. Although there were no significant differences in bow pressure values between the experienced players and beginners, the range was larger for beginners, and both the maximum and minimum values ($p = 0.059$) were smaller for beginners. Seven of the eight experienced players had a minimum bow pressure of 0.5 N or more. Generally, the weight of a bow is around 60 g, which translates to approximately 0.59 N in Newton units. In this study, the experienced players usually kept bow pressure higher than or as high as the bow weight.

Next, we examined the relationship between bow position and bow pressure. Excluding the specified range of bow position, we divided the entire length of the bow into three sections: the frog, middle, and tip. Fig. 4 shows the bow pressure during down-bows and up-bows for experienced players and beginners. Bow pressure was normalized based on each participant's maximum and minimum values. The Wilcoxon signed-rank test was conducted at a significance level of 5%, with the Bonferroni correction applied for multiple comparisons of bow pressures between the three sections. During the down-bows, experienced players had significantly lower bow pressure in the middle section compared to the frog but no significant differences between the middle and tip or the frog and tip. These results suggested that the experienced players applied the necessary force at the bow frog and tried to keep consistent bow pressure, even at the tip during down-bows. We did not observe the same trend in the up-bows of the experienced players, but there was no significant difference between the frog and middle sections compared to the beginners. These results indicated that the experienced players tried to maintain consistent bow pressure from the tip toward the frog once the bow stroke started. On the other hand, beginners had significantly lower bow pressure at the tip compared to the frog, regardless of the direction of the bow stroke. As the tip is farther from the hand, the pressure could have weakened at the tip compared to that of the frog.

Next, we examined the relationship between bow pressure and bow speed. We normalized bow pressure and speed for each participant. We divided bow pressure into three stages based on the magnitude. We conducted the Wilcoxon signed-rank test at a significance level of 5%, with Bonferroni correction applied for multiple comparisons of bow speeds between the three stages. As shown in Fig. 5, we observed no significant differences in bow speeds between the three stages of bow pressure for either the experienced players or the beginners. However, the beginners tended to have slower bow speeds when bow pressure was high, whereas the experienced players tended to have slower bow speeds when bow pressure was low. According to Blanco et al. [6], when the players apply more force on a violin string, they need to move the bow faster to avoid the violin sound being scratchy, whereas when they apply less force on the string, they need to move the bow slower to prevent the sound being too abrasive. In this experiment, the experienced players adjusted the relationship between bow pressure and bow speed to maintain sound quality.

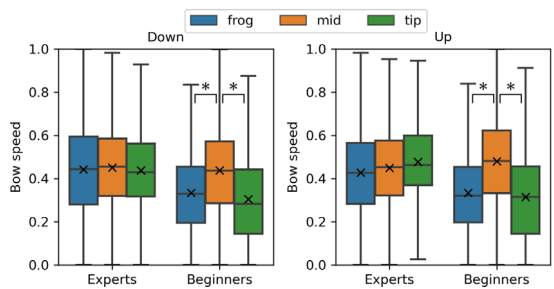

Fig.6 Relationship between bow position and speed

### 3.2 Bow position and bow speed

We also examined the relationship between bow position and bow speed. Fig. 6 shows the bow speed during down-bows and up-bows for experienced players and beginners. Bow speed was normalized based on each participant's maximum and minimum values. We conducted the Wilcoxon signed-rank test at a significance level of 5%, with Bonferroni correction applied for multiple comparisons of bow speeds between three sections. Regardless of the direction of the bow stroke, we observed no significant differences in bow speed among the three sections for the experienced players. However, the beginners exhibited significant differences in bow speed between the frog and middle sections, as well as between the middle and tip sections. The beginners had faster bow speeds in the middle section and slower speeds at the frog and tip. This could be due to the difficulty in managing the pace of bowing to match the tempo.

Compared to the beginners' graphs shown in Fig. 3, where the peaks at the tip and valleys at the frog indicate the direction change of the bow, the experienced players' graphs had sharper peaks and valleys. This suggested that the beginners took longer to change their stroke direction. To verify this, we detected peaks from the time-series data of the bow position and calculated the curvature at those points. We tested the curvatures at the frog and the tip of the experienced players and the beginners using the Brunner-Munzel test at a significance level of 5%. The curvature was significantly larger at both the tip and the frog for experienced players. This result indicates that the experienced players could transition the direction of the bow more smoothly and quickly.

### 4 Conclusions

This study measured and analyzed the bowing movements of violin beginners and experienced players. The results showed that the experienced players consistently applied a higher minimum bow pressure and, particularly during down-bows, applied higher bow pressure only at the beginning of the stroke, then quickly reduced it while keeping the bow pressure steady until reaching the tip. In addition, the experienced players exhibited faster bow direction changes and maintained a consistent bow speed, regardless of the bow position. Developing a feedback system based on experienced players' bowing performance and guiding beginners to match this performance will improve the efficiency of their practice.

ACKNOWLEDGEMENT

Part of this work was supported by JST PRESTO (Grant Number JPMJPR2134).